# Investigation of Mode Coupling in Normal Dispersion Silicon Nitride Microresonators for Kerr Frequency Comb Generation

Yang Liu[1], Yi Xuan[1,2], Xiaoxiao Xue[1], Pei-Hsun Wang[1], Steven Chen[1], Andrew J. Metcalf[1], Jian Wang[1,2], Daniel E. Leaird[1], Minghao Qi[1,2], and Andrew M. Weiner[1,2*]

[1]*School of Electrical and Computer Engineering, Purdue University, 465 Northwestern Avenue, West Lafayette, IN 47907-2035*
[2]*Birck Nanotechnology Center, Purdue University, 1205 West State Street, West Lafayette, Indiana 47907, USA*
*\*Corresponding author: amw@purdue.edu*

**Kerr frequency combs generated from microresonators are the subject of intense study. Most research employs microresonators with anomalous dispersion, for which modulation instability is believed to play a key role in initiation of the comb. Comb generation in normal dispersion microresonators has also been reported but is less well understood. Here we report a detailed investigation of few-moded, normal dispersion silicon nitride microresonators, showing that mode coupling can strongly modify the local dispersion, even changing its sign. We demonstrate a link between mode coupling and initiation of comb generation by showing experimentally, for the first time to our knowledge, pinning of one of the initial comb sidebands near a mode crossing frequency. Associated with this route to comb formation, we observe direct generation of coherent, bandwidth-limited pulses at repetition rates down to 75 GHz, without the need to first pass through a chaotic state.**

Recently high quality factor (Q) microresonators have been intensively investigated for optical comb generation. Both whispering gallery mode resonators employing tapered fiber coupling and chip-scale microresonators employing monolithically fabricated coupling waveguides are popular. Tuning a continuous-wave (CW) laser into resonance leads to build-up of the intracavity power and enables additional cavity modes to oscillate through cascaded four-wave mixing (FWM) [1-15]. Modulational instability (MI) of the CW pump mode is commonly cited as an important mechanism for comb generation [16-18]. According both to experiment and to theoretical analysis, comb generation preferably occurs in resonators with anomalous dispersion. However, comb generation in resonators characterized with normal dispersion has also been observed experimentally [5, 8, 19-24]. Several models have been proposed to describe this phenomenon. Although MI gain is missing in fibers or waveguides with normal dispersion, when it comes to resonators, the detuning provides an extra degree of freedom which enables MI to take place in the normal dispersion regime, hence providing a route to comb generation [16, 18, 25]. However, this mechanism requires either a precise relationship between detuning and pump power, making it difficult to realize practically, or hard excitation, a nonadiabatic process under which pump photons must be initially present in the resonator [17].

Mode coupling has also been suggested as a mechanism enabling comb generation in resonators with normal dispersion [26]. When resonances corresponding to different families of transverse modes approach each other in frequency, they may interact due to imperfections in the resonator. The theory of mode coupling in resonators has been well-established [27], and frequency shifts and avoided crossings have been observed [28-33]. In the anomalous dispersion regime, mode coupling has been reported to affect the bandwidth scaling of frequency combs [34] and the process of soliton formation [35]. However, in these cases anomalous dispersion is still considered to be the determining factor for comb generation; mode coupling is considered to be detrimental, inhibiting the formation of solitons and limiting comb bandwidth. In the normal dispersion regime, measurements have been performed with $CaF_2$ whispering gallery mode resonators [26]. The experiments demonstrate strong local frequency shifts that are attributed to mode interactions and show a correlation between the presence of such local frequency shifts and the ability to generate combs in these normal dispersion resonators. Significant changes in comb spectra have been observed when pumping different longitudinal modes spaced by only a few free spectral ranges (FSR), both in normal dispersion silicon nitride microring resonators [5] and in the whispering gallery mode resonators of [26]; in the latter case, such effects were specifically attributed to mode interactions.

In the current report, we perform comb-assisted precision spectroscopy measurements [36] of few-moded silicon nitride microresonators in the normal dispersion regime over frequency ranges spanning dozens of FSRs. As a result we are able to clearly map out mode interactions and obtain plots of resonant frequencies exhibiting strong avoided crossings closely analogous to those that occur for quantum mechanical energy surfaces [37-39]. The frequency shifts affecting a series of resonances from both mode families can lead to a strong change in local dispersion, even changing its sign. We provide clear experimental evidence that this mode coupling plays a major role in the comb generation process for our normal dispersion resonators by showing experimentally, for the first time to our knowledge, that the location of one of the two initial sidebands at the onset of comb generation is "pinned" at a mode crossing frequency, even as the pump wavelength is changed substantially[40].

These effects allow us to realize a "Type I" comb [5], also termed a natively mode-spaced (NMS) comb [41] in a resonator with FSR slightly under 75 GHz. In such a comb the initial sidebands are generated via a soft excitation mechanism and are spaced one FSR from the pump; the comb exhibits low noise and high coherence immediately upon generation [5, 20-22, 41]. We also find that the Type-I comb as generated here corresponds directly to a train of bandwidth-limited pulses. This is in sharp contrast to "Type II" combs [5] (also termed multiple mode-spaced (MMS) combs [41]) for which the initial sidebands are separated from the pump by several FSR, after which additional, more closely sidebands are generated (usually with increasing intracavity power) to arrive at single FSR spacing. Such Type II combs exhibit poor coherence and high noise [5, 21, 41, 42]. Mode-locking transitions in which Type II combs switch into a coherent, low noise state have been observed experimentally and studied theoretically [17, 23-25, 35, 41, 43-54]. However, these methods require careful and sometimes complex tuning of the pump frequency or power [35]; the mode-locking transition is often difficult to achieve and until very recently has not been observed in normal dispersion microresonators. Our recent demonstration of dark soliton formation in resonators with normal dispersion is linked to a mode-locking transition [23], but the waveforms generated are quite distinct from the bandwidth-limited pulses reported here.

The field in the microresonators can be expressed using the following mode coupling equations:

$$\frac{dE_1}{dt} = \left(-\frac{1}{\tau_{o1}} - \frac{1}{\tau_{e1}} - j\delta_1\right)E_1 + j\kappa_{12}E_2 + \sqrt{\frac{2}{\tau_{e1}}}E_0 \quad (1a)$$

$$\frac{dE_2}{dt} = \left(-\frac{1}{\tau_{o2}} - \frac{1}{\tau_{e2}} - j\delta_2\right)E_2 + j\kappa_{21}E_1 + \sqrt{\frac{2}{\tau_{e2}}}E_0 \quad (1b)$$

Here $E_1$ and $E_2$ are the intracavity fields for mode 1 and 2 respectively, $1/\tau_{o1}$ and $1/\tau_{o2}$ are decay rates due to the intrinsic loss for both modes while $1/\tau_{e1}$ and $1/\tau_{e2}$ are coupling rates between the resonator and the bus waveguides. $\delta_1 = \omega - \omega_1$ and $\delta_2 = \omega - \omega_2$ are the frequency detunings where $\omega_1$ and $\omega_2$ are the resonant frequencies. $\kappa_{12} = \kappa_{21}^* = \kappa$ are mode coupling coefficients. We can simulate mode interaction effects using the mode coupling equations. Two modes are assumed working close to 1550nm in the resonator with $\tau_{o1} = 2.99 \times 10^{-9}s$, $\tau_{e1} = 3.65 \times 10^{-9}s$ and $\tau_{o1} = 1.25 \times 10^{-9}s$, $\tau_{e1} = 2.41 \times 10^{-9}s$, respectively. This corresponds to two modes working in the under-coupling regime: mode 1 with loaded Q of $10^6$, intrinsic Q of $1.82 \times 10^6$ and extinction ratio of 10 dB; mode 2 with loaded Q of $5 \times 10^5$, intrinsic Q of $7.6 \times 10^5$ and extinction ratio of 5 dB, respectively. We solve eq. (1) and plot the resulting transmission spectra for different separations (over the range -5 GHz to 5 GHz) between the resonances of the two modes. Without mode coupling ($\kappa = 0$, Fig. 1(a)), the resonances approach and cross each other at a constant rate. However, with mode coupling turned on ($\kappa = 5 \times 10^9 s^{-1}$ compared with $1/\tau_{e1} = 2.74 \times 10^8 s^{-1}$, Fig. 1(b)), the dips in transmission are shifted in frequency, resulting in an avoided crossing. The mode interactions also lead to significant changes in the extinction ratios and linewidths of the resonant features. Similar effects are observed in our experiments, as we relate below.

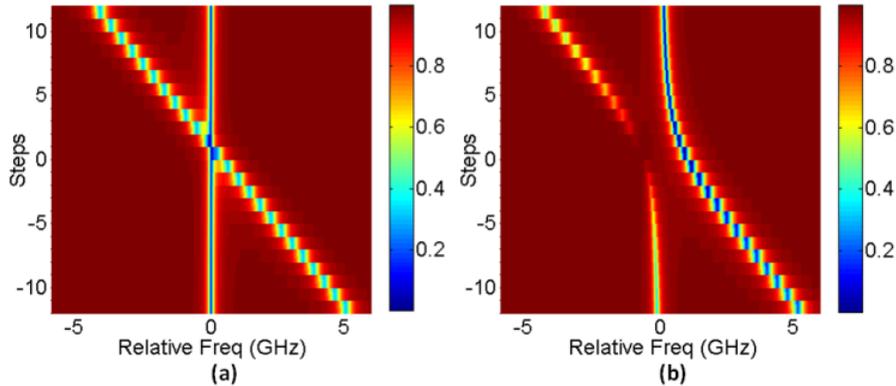

Fig. 1 Numerical investigation of mode coupling effect when the resonances of the two modes get close and cross each other (a) no mode coupling (b) with mode coupling.

Our experiments utilize silicon nitride resonators fabricated to have 2 μm × 550 nm waveguide cross-section. According to simulation for two transverse electric (TE) modes, TE$_1$ and TE$_2$, these waveguides are clearly in the normal dispersion regime with D~ $-156$ ps/nm·km and D~ $-160$ ps/nm·km respectively [20]. We first study a resonator with a total path length of 5.92 mm which corresponds to a FSR slightly under 25 GHz. Similar to [14], to avoid the stitching error we introduce a finger-shaped structure for the resonator so that it can fit in a single field of our electron beam lithography tool. Figure 2(a) shows a microscope image of the microresonator. The light is coupled both in and out through lensed fibers which are positioned in U-grooves to improve stability when working at high power [20]. Fiber-to-fiber coupling loss is ~5 dB. The measured transmission spectrum, showing resonances throughout the lightwave C band, is given in Fig. 2(b). If we zoom in the transmission spectrum as shown in the inset, resonances of 2 transverse mode families with different depth can be observed. The loaded Q factors at the frequencies shown are ca. $1\times10^6$ (intrinsic Q~$1.7\times10^6$) and $0.3\times10^6$ (intrinsic Q~$0.35\times10^6$) for modes 1 and 2, respectively.

We use the frequency-comb-assisted spectroscopy method of [36] to accurately determine the resonance positions and compute the changes in FSR with wavelength to estimate the dispersion for TE modes. The measured FSRs are given in Fig. 3(a). The FSRs for the two modes are around 24.8 GHz and 24.4 GHz, respectively. Both modes are fitted with our simulated dispersion showing good accordance which confirms that the resonator is in the normal dispersion regime for TE modes with $\frac{D_2}{2\pi} \approx 474\ kHz$, where $\frac{D_2}{2\pi}$ denotes the difference in FSR for adjacent resonances which can be expressed as:

$$\frac{D_2}{2\pi} = \lambda^2 \cdot n_{eff} \cdot FSR^2 \cdot D \qquad (2)$$

However, at several wavelengths for which the resonances associated with the two transverse modes are closely spaced (1532 nm, 1542 nm and 1562 nm), the FSRs of the two modes change significantly, such that their FSRs become more similar. In these cases we clearly observe that the mode coupling results in a major modification to the local dispersion, even changing the sign of dispersion in some wavelength regions. To take a closer look at this phenomenon, in Fig. 3(b) we plot the transmission spectrum in the vicinity of the mode crossing regime near 1542 nm. To visualize the data in a form analogous to the simulations of Fig. 1, we vertically align different pieces of the transmission spectrum separated by a constant 24.82 GHz increment (the nominal FSR of the higher Q mode around 1542 nm). Since the average dispersion contributes a change in FSR below 15 MHz in the range plotted without mode coupling, one of the modes should appear as very nearly a vertical line, while the other should appear as a tilted line due to the difference in FSRs. However in Fig. 3(b), we observe that the curves bend as they approach each other, resulting in an avoided crossing, similar to the simulation results of Fig. 1(b). Changes in the extinction ratio of the resonances are also clearly evident in the mode interaction region. These data provide detailed and compelling evidence of strong mode coupling effects on the linear spectrum.

A different case for the mode crossing is observed around 1552 nm. Here there is no obvious change in FSR around the wavelength where the resonances of these modes get close. The aligned resonance pairs are shown in Fig. 3(c). The picture resembles the case shown in Fig. 1(a), where no mode coupling is assumed. However, if we zoom-in on the data, we can again see slight shifts in the positions of the resonances when they are close enough. In this case mode coupling effects are present but weak.

In comb generation experiments, we pumped the micro-resonator with a single CW input at 1.75W (this is the value prior to the chip, without accounting for coupling loss) tuned to different resonances of the high-Q mode family and recorded the comb spectra. The results are given in Fig. 4. In Fig. 4(a) we pump at 28 different resonances between 1554 and 1560 nm. The frequency spacing of the comb varies from 33 FSR for pumping at 1554 nm to 7 FSR for pumping at 1559.4 nm. We observe that the nearest long wavelength sideband remains anchored at approximately 1560.5 nm, very close to the ~1562 nm mode interaction feature. With the pump shifted by a total of 694 GHz (27 FSRs), we find that the long wavelength sideband varies by no more than ±25GHz (±1FSR). Meanwhile the short wavelength sideband varies at twice the rate of the pump tuning, for a total frequency variation of ~1.3 THz. Similar behavior is observed when we pump between 1546nm to 1549.5 nm. As shown in Fig. 4(b), one of the sidebands is anchored near 1550.5 nm, close to the weaker 1552 nm mode interaction feature. In this case comb generation is missing for some pump wavelengths, which may be the result of the weak coupling strength. The observed pinning of one of the initial sidebands very close to a mode interaction feature clearly suggests that mode coupling is a major factor in comb generation in this normal dispersion microresonator.

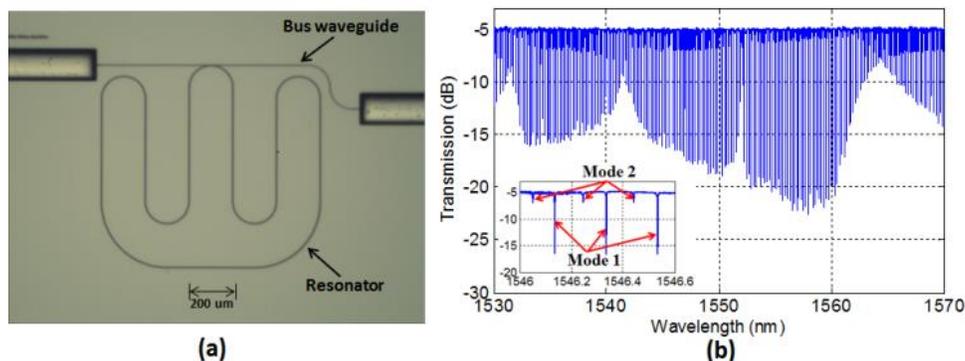

Fig. 2. (a) Microscope image of the silicon nitride resonator with path length of 5.92 mm. (b) Measured transmission spectrum of the resonator. Inset is the zoom-in transmission spectrum showing resonances from different transverse mode families.

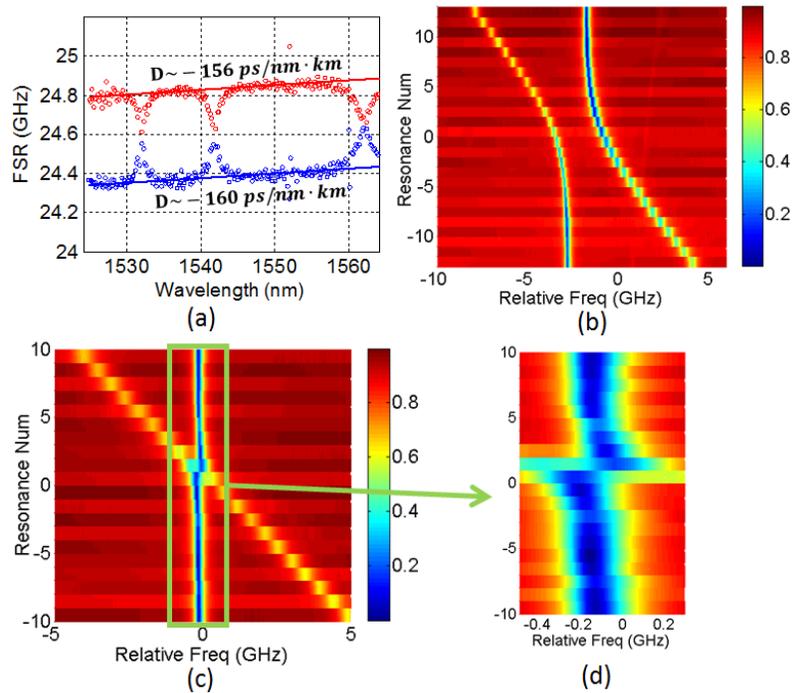

Fig. 3. (a) Measured FSR versus optical wavelength for two TE modes, plotted in red and blue and fitted with $D \sim -156\,ps/nm \cdot km$ and $D \sim -160\,ps/nm \cdot km$ respectively. (b-d) Aligned resonances with fixed increment showing the mode coupling with different coupling strength (b) Strong coupling case centered at 1542 nm with clear avoided crossings (c) Weak coupling case centered at 1552 nm (d) Zoomed-in view of the weak coupling case

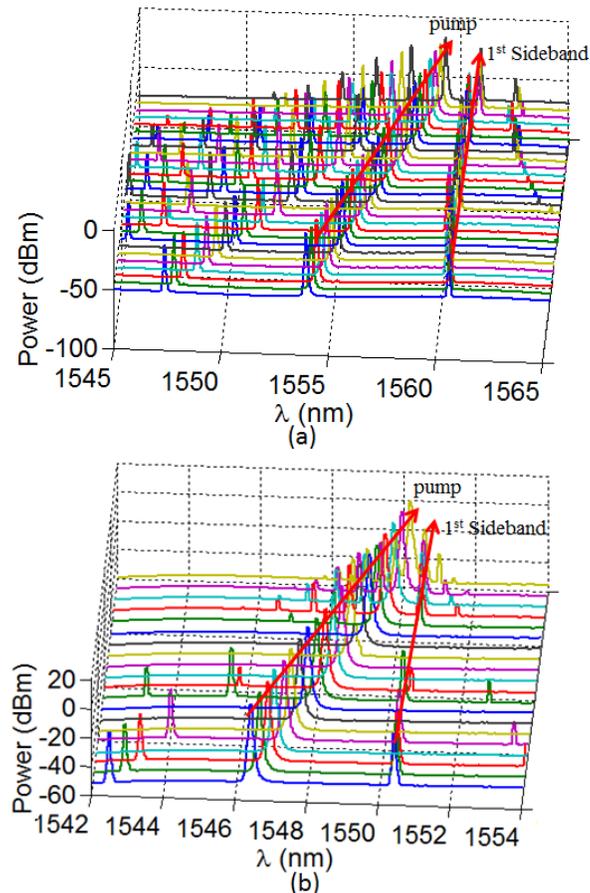

Fig. 4 Comb generation at different pump wavelength with one of the 1st sidebands kept at an approximately constant location (a) around 1560 nm and (b) around 1550 nm

According to the anomalous dispersion analysis of Ref. [41], increasingly large dispersion is needed to generate "NMS" (Type I) combs as the resonator FSR decreases. Physically the increased dispersion brings the MI gain peak closer to the pump. In order to reduce the gain peak to single FSR frequency offset, it was shown that the $D_2$ parameter should be made close to the resonance width (200 MHz for silicon nitride resonators with Q~$10^6$). For example, for a resonator with FSR~100 GHz, the dispersion required for the generation of a Type I comb would be $D \approx 4.2 \times 10^3 ps \cdot nm^{-1} \cdot km^{-1}$. Furthermore, according to Eq.(2), the dispersion required grows quadratically as the resonator size is further increased (required D grows as inverse square of the FSR). Such dispersions are generally too large to achieve practically; perhaps as a result, no observation of Type I comb generation in sub-100 GHz silicon nitride resonators has been reported. However, mode coupling can dramatically change the local dispersion, both increasing its magnitude and changing its sign. Generation of Type I combs from large whispering gallery mode (WGM) resonators has previously been reported and attributed to mode coupling [25, 26]. In our experiments we have observed Type I comb generation in SiN for a resonator with a FSR slightly below 75 GHz. We have not yet obtained a NMS comb from resonators with even smaller FSR. However, as shown in Fig. 4(b), a comb with 2-FSR separation is observed when the 25 GHz FSR resonator is pumped at 1550.41 nm. This means that the 1st sideband is less than 50 GHz from the pump, which is still very difficult to achieve without mode coupling effects.

The fabricated ~75 GHz FSR resonator is shown in Fig. 5(a). Unlike the 25-GHz resonator discussed earlier, it has a drop-port design which has been observed to reduce the power difference between the pump and adjacent comb lines, yielding a smoother comb spectrum without the usual strong pump background [20]. Using the frequency-comb-assisted spectroscopy method, two mode families with FSRs around 74.7 GHz and 72 GHz are observed. The two families of resonances approach each other around 1563 nm. Different sections of the transmission spectrum are aligned in a similar fashion as for Fig. 3(b-d) and plotted in Fig. 5(b). An avoided crossing evidencing mode coupling is clearly observed. The comb results are shown in Fig. 5(c) for pumping between 1554.71 nm and 1566.59 nm. Again one of the 1st sidebands is "pinned" near the mode crossing wavelength. Although the first sideband has a 13-FSR separation when pumping at 1555 nm, a Type I comb can be generated for pumping at 1562.62 nm, 1563.22 nm, 1563.81 and 1564.43nm (as shown in the circled area). We note that in this case, pinning of the 1st sideband is observed for pumping at either the blue side or the red side of the mode crossing area.

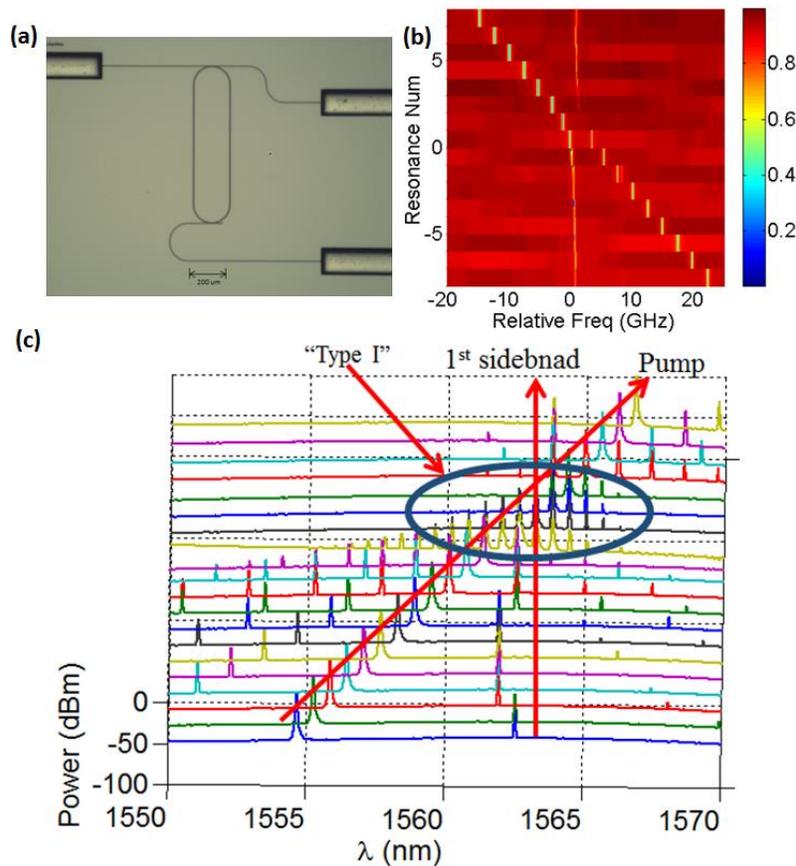

Fig. 5 (a) Microscope image of the silicon nitride resonator with path length of 1.97 mm. (b) Aligned resonances centered at 1563 nm with fixed increment showing the mode coupling. (c) Comb generation at different pump wavelength; one of the 1st sidebands remains close to 1563 nm

As an example, pumping at 1562.62 nm with 1.6W input, more than 20 comb lines with 1 FSR separation are generated. The spectrum observed at the drop port is shown in Fig. 6(a). Fifteen of the lines are selected by a bandpass filter and amplified in an EDFA; the resulting spectrum is shown in Fig. 6(b). The amplified and filtered comb is directed to an intensity autocorrelator based on second harmonic generation in a noncollinear geometry. A length of dispersion compensating fiber (DCF) is used to achieve dispersion compensation of the entire fiber link (including the EDFA) connecting the SiN chip to the autocorrelator. The length of DCF was adjusted by injecting a short pulse laser from a passively mode-locked fiber laser into the front end of the fiber link and minimizing its autocorrelation width. The autocorrelation trace measured for the comb is plotted in Fig. 6(c). Also plotted is the autocorrelation of the ideal bandwidth-limited pulse, calculated from the spectrum of Fig. 6(b) with the assumption of flat phase. Clearly the generated pulses, with estimated duration of 2.7 ps FWHM, are very close to bandwidth-limited. We have also used a photodetector and spectrum analyzer to look at the low frequency intensity noise of the comb (measurement bandwidth: ~500 MHz). The intensity noise is below the background level of our measurement setup. Similar low noise, bandwidth-limited pulse generation is observed for the "Type I" combs generated via pumping at other resonances of this same resonator. These data demonstrate that the Type I combs reported here are generated *directly* in a mode-locked state featuring low noise, high coherence, and bandwidth-limited temporal profile, though with a limited number of comb lines.

Our group has previously reported direct Type I generation, with behavior similar to that shown in Fig. 6, from a smaller, normal dispersion SiN microresonator with ~230 GHz FSR [20]. Although we speculated that mode interactions may have played a role in allowing comb generation, as pointed out theoretically in [26], we were unable to present data to support this speculation. Based on the insight developed in the current paper, we decided to reexamine our data from the device of [20]. Figure 7 shows the comb spectra obtained for pump powers just above threshold, plotted in the same fashion as Figs. 4 and 5(c). This format clearly shows pinning of one of the initial sidebands, revealing what we now understand to be a signature of comb initiation through mode coupling.

The pathway to coherent pulse generation reported here is clearly distinct from that observed in [23, 24, 35, 42], which attain broader comb bandwidths but need to navigate through a chaotic state before arriving at a transition to mode-locking. Although a theoretical explanation for this phenomenon is still not fully available, our simulation using the Lugiato-Lefever (L-L) equation [23, 45, 46, 55] shows that stable, coherent Type I combs can be induced in normal dispersion silicon nitride resonators by introducing a phase shift term (as suggested by Ref. [26] to model the effect of mode interaction) to a single resonance adjacent to the pump resonance. However, the Type I combs in our simulations, though coherent, do not correspond to bandwidth-limited pulses; instead the comb field is phase modulated and temporally broadened. This discrepancy suggests some factors may be missing in the model. One possibility is that instead of adding a phase shift to a single resonance, phase shifts should be assigned to a group of resonances distributed around the mode interaction region. Another, more radical idea is that a heretofore unidentified nonlinear amplitude modulation mechanism may be present, as suggested briefly in Ref. [20, 41]. Mode interactions may contribute to such a mechanism, since a superposition of transverse modes leads to a longitudinally modulated spatial profile which may either increase or decrease overlap with waveguide imperfections. Nonlinearity could shift such spatial profiles, under appropriate circumstances reducing loss, analogous to decreased loss through nonlinear lensing in Kerr lens mode-locked lasers. Another possibility is that wavelength dependent Q introduces a spectral filtering effect which contributes to shaping the time domain field, as suggested in Ref. [24].

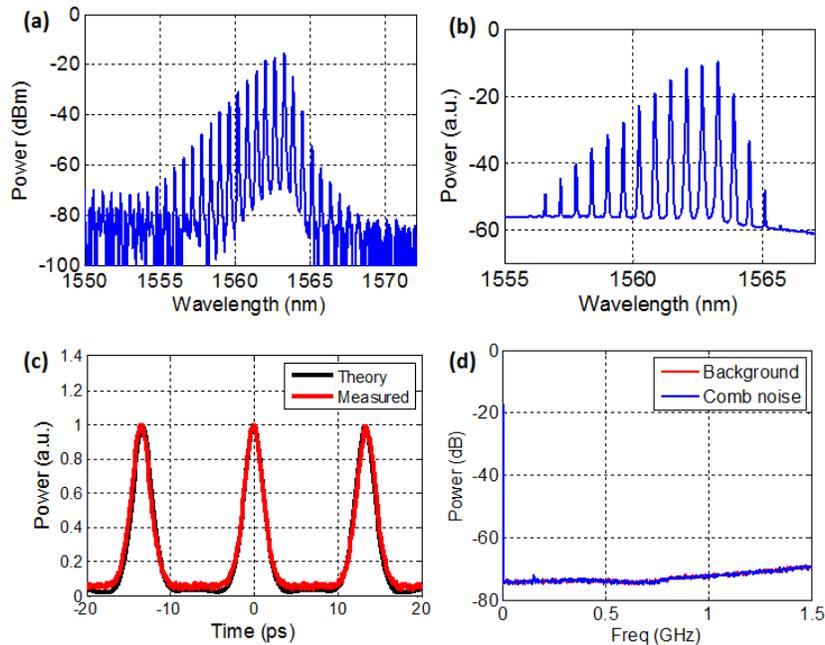

Fig. 6 Generation of coherent mode-locked Type I comb due to the mode coupling (a) Generated comb spectrum at the drop port (b) 15 lines are selected, amplified and filtered for time-domain characterization. (c) Autocorrelation of time domain pulse compared with that of theoretical bandwidth-limited pulse, showing good coherence and mode locking behavior. (d) Intensity noise compared with the measurement system noise floor

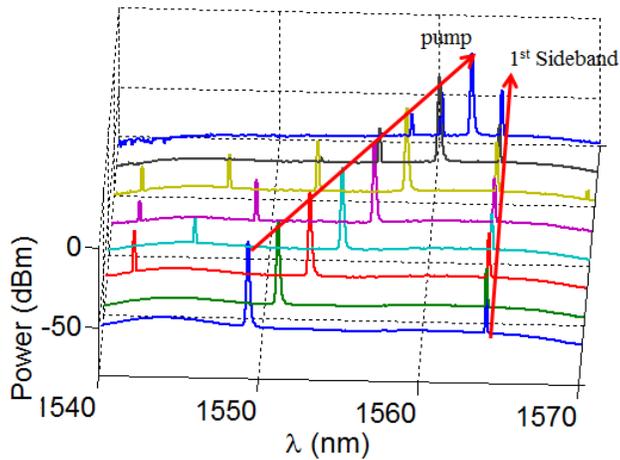

Fig. 7 Observation of "Pinned" 1st sidebands for the resonator with passively mode locked "Type I" comb as discussed in Ref. [20]

Finally, we note that previous studies have associated mode coupling with asymmetric comb spectra [19, 26]. However, mode interaction spectra such as those in Figs. 3 and 5 were neither reported nor registered with the generated combs. In our experiments asymmetric spectra were observed for both resonators studied; Fig. 8 shows four examples with the mode interaction region identified through the pinned 1st sideband. For the resonator with ~25 GHz FSR, the separations between the pump and the 1st sideband are 15 FSR and 2 FSR in Figs. 8(a) and 8(b), respectively; for the ~75 GHz FSR resonator, the pump position is changed from the short wavelength side of the coupling region (Fig. 8(c)) to the long wavelength side (Fig. 8(d)). In each case the 1st sideband "pinned" close to the mode crossing has higher power than the 1st sideband on the other side of the pump. However, fewer comb lines are generated on the side of the pump corresponding to the pinned sideband. We may understand this behavior by noting that flattened dispersion is favorable for broadband comb generation [13, 46, 56]. In our experiments mode coupling modifies the local dispersion, allowing MI gain for initiation of comb generation but at the same time giving rise to significant higher order dispersion that limits growth of the comb bandwidth on the mode interaction side.

In summary, we have demonstrated what we believe to be conclusive evidence of the impact of mode coupling on initiation of comb generation in normal dispersion silicon nitride microresonators. We have also demonstrated mode- coupling-assisted "Type I" comb generation resulting in direct generation of bandwidth-limited pulses, without the need to navigate through a chaotic state.

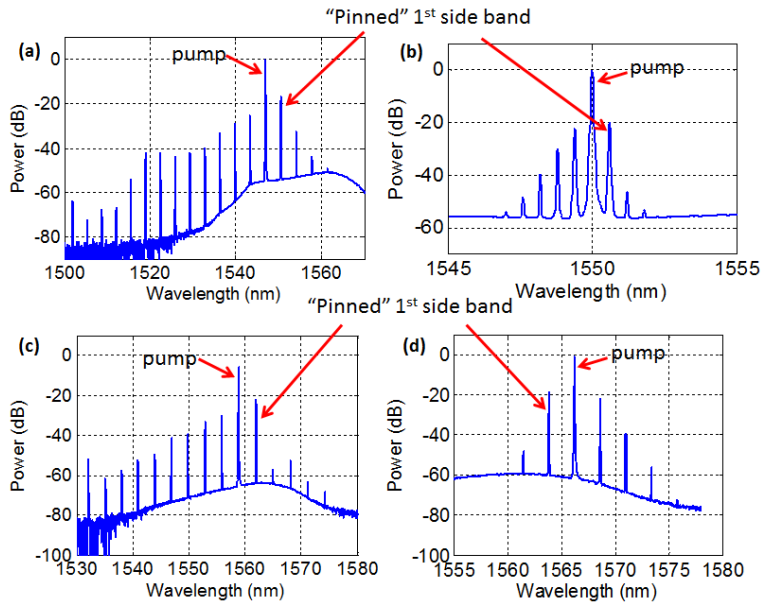

Fig.8 Asymmetric comb spectra. (a) and (b) comb generation using resonators with ~25 GHz FSR, with "Pinned" sideband close to 1551nm; (c) and (d) comb generation using resonators with ~75 GHz FSR, with "Pinned" sideband close to 1563 nm.

**Acknowledgements**

This work was supported in part by the National Science Foundation under grant ECCS-1102110, by the Air Force Office of Scientific Research under grant FA9550-12-1-0236, and by the DARPA PULSE program through grant W31P40-13-1-0018 from AMRDEC.